\documentclass[a4paper,12pt]{article}
\usepackage{amsmath}
\usepackage{subcaption}
\usepackage{color}
\usepackage{graphicx,psfrag,epsf}
\usepackage{enumerate}
\usepackage{enumitem}

\usepackage{natbib}

\usepackage{DNstyle}
\usepackage{url} % not crucial - just used below for the URL 

\newcommand{\blind}{1}
\date{}
\begin{document}

	\def\spacingset#1{\renewcommand{\baselinestretch}%
		{#1}\small\normalsize} \spacingset{1}

	%%%%%%%%%%%%%%%%%%%%%%%%%%%%%%%%%%%%%%%%%%%%%%%%%%%%%%%%%%%%%%%%%%%%%%%%%%%%%%
	
	\if1\blind
	{
		\title{\bf A novel calibration framework for survival analysis when a binary covariate is measured at sparse time points}
		\author{Daniel Nevo\thanks{ \footnotesize \noindent Departments of Biostatistics and Epidemiology, Harvard T.H. Chan School of Public Health.  DN and MW are co-corresponding authors. \hfill \break
				$\dagger$ Department of Oncologic Pathology, 
				Dana-Farber Cancer Institute and Harvard Medical School 
				\hfill \break
				$\ddagger$ Department of Epidemiology, Harvard T.H. Chan School of Public Health \&
				Program in MPE Molecular Pathological Epidemiology, Department of Pathology,
				Brigham and Women`s Hospital and Harvard Medical School
				\hfill \break
				$\mathsection$ Channing Division of Network \& Medicine, Department of Medicine, Brigham and Women's Hospital and Harvard Medical School
				\hfill \break
				This work was supported by U.S. National Institutes of Health (NIH) grants R35 CA197735, R01 CA151993, UM1 CA186107, P01 CA87969, P01 CA55075, UM1 CA167552,  U01 CA167552, 	R01 CA118553, and R01 CA169141  }\hspace{.2cm},
			Tsuyoshi Hamada$^\dagger$,
			Shuji Ogino$^{\dagger\ddagger}$
			and 
			Molin Wang$^{*, \mathsection}$}
		\maketitle
} \fi
	
	\if0\blind
	{
		\bigskip
		\bigskip
		\bigskip
		\begin{center}
			{\LARGE\bf A novel calibration framework for survival analysis when a binary covariate is measured at sparse time points}
		\end{center}
		\medskip
	} \fi
	
	\bigskip
\begin{abstract}
%	A common goal in clinical studies is to evaluate the association of a time-dependent binary treatment or exposure with time-to-event.  Recent group of impactful colorectal cancer studies targeted the association between aspirin-taking, a binary time-dependent covariate, and survival following colorectal cancer diagnosis. Due to surgery, aspirin-taking value is assumed zero at the beginning of the study and may change its value to one at some time point. Estimating these associations from cohort studies is complicated by having only intermittent measurements on aspirin-taking.   The proportional hazard model cannot be fitted since the value of the covariate is observed only in certain time points, and thus its exact value is unknown for some participants in each risk set. Commonly-used n\"aive methods  are potentially biased, especially when the covariate is measured at sparse time points. We present a class of calibration models for the distribution of the time of status change of the binary covariate. Estimates obtained from these models are then incorporated into the proportional hazard partial likelihood in a natural way. We develop nonparametric, semiparametric and parametric models, and derive asymptotic theory for the methods we implement in the aspirin and colorectal cancer data. Our methodology allows for inclusion of additional baseline variables affecting the  status change time of the binary covariate.  We further develop a risk-set calibration approach for covariates with strong effect on the time-to-event. 
The goals in clinical and cohort studies often include evaluation of the association of a time-dependent binary treatment or exposure with a survival outcome.  Recently, several impactful studies targeted the association between aspirin-taking and survival following colorectal cancer diagnosis. Due to surgery, aspirin-taking value is zero at baseline and may change its value to one at some time point. Estimating this association is complicated by having only intermittent measurements on aspirin-taking.   N\"aive, commonly-used,  methods can lead to substantial bias. We present a class of calibration models for the distribution of the time of status change of the binary covariate. Estimates obtained from these models are then incorporated into the proportional hazard partial likelihood in a natural way. We develop nonparametric, semiparametric and parametric calibration models, and derive asymptotic theory for the methods that we implement in the aspirin and colorectal cancer study. Our methodology allows to include  additional baseline variables in the calibration models for the status change time of the binary covariate.  We further develop a risk-set calibration approach that is more useful in settings in which the association between the binary covariate and survival is strong.
\end{abstract}
	
	\noindent%
\scriptsize{	{\it Keywords:}  Proportional Hazard; Interval Censoring; Missing Data;  Last-value-carried-forward; Imputation.}
	\vfill
	
	\newpage
	\spacingset{1.45} % DON'T change the spacing!
\section{Introduction}
\label{Sec:Intro}
One benefit of the Cox proportional hazards (PH) model  for analysis of time-to-event data \citep{cox1972regression} is the simplicity of including time-dependent covariates, while preserving desirable theoretical properties. Classical methods assume that time-dependent covariates are measured continuously. However, in practice,  they are often measured intermittently,  leading to bias in effect estimates if treated n\"aively \citep{andersen2003attenuation,cao2015analysis}.   We consider a time-dependent binary covariate having zero value at baseline, that may change its value to one at some point, and once a change has occurred, the covariate retains this value for the rest of the follow-up time. Real-life scenarios of this nature are widespread, including the onset of a irreversible medical condition \citep[e.g., HIV infection, ][]{langohr2004parametric}  or a treatment with a constant effect that is administrated in a different time for each patient \citep{austin2012generating}. For example, \cite{goggins1999applying} described  data arising from a clinical trial where the goal was to study the effect of Cytomegalovirus (CMV) shedding on risk of developing active CMV disease.

In the problems motivating this paper,   the researchers  were interested in  the association between  regularly taking aspirin, henceforth simply referred to as aspirin-taking,  and the survival time  following the diagnosis of colorectal cancer (CRC).  A series of studies published in leading clinical journals \citep{chan2009aspirin,liao2012aspirin,hamada2017aspirin}  provided evidence for differential association between post-diagnosis aspirin-taking and the survival time  according to tumor subtype classification, which is a baseline variable coded as a categorical variable.   The main dataset was obtained from two cohort studies: the Nurses' Health Study (NHS) and the Health Professionals Follow-Up Study (HPFS).  Following their enrollment to the studies, participants have been receiving questionnaires biennially and answering questions about life-style and other participants' characteristics.  Patients  diagnosed with colorectal cancer who had been taking aspirin typically stop taking aspirin at the time of diagnosis as part of their preparations to surgery. 
 Post-diagnosis aspirin-taking is a time-dependent binary covariate, and the data about it is incomplete. The time a participant started to take aspirin is  known to lie within the interval between the time of the last questionnaire  answered  as no  aspirin-taking and the  time of the first questionnaire  answered as aspirin-taking.   A popular strategy is to replace unknown values of the time-dependent covariate with the last available value.  Another alternative is to impute the change-time, the time of change in the covariate value  from 0 to 1, using  the middle point of the observed interval. 

There is little existing literature relevant to this statistical challenge. For time-to-event data, \cite{andersen2003attenuation} studied the attenuation in effect estimates caused by having infrequently measured covariates and, more recently, \cite{cao2015analysis} developed  kernel-based weighted score methods.  \cite{goggins1999applying} considered the problem of interval-censored change-time of a binary covariate  and proposed an EM algorithm to estimate the association between the binary covariate and a survival time. \cite{langohr2004parametric} developed a parametric log-linear survival model for the interval-censored change-time problem. None of these methods allow to use covariates that are informative about the change-time of the binary covariate.

In this paper,  we propose a novel analysis framework for the problem of interval-censored change-time and suggest a two-stage approach. In the first stage, we fit a model  for the status-changing time of the binary covariate. This model may be nonparametric, semiparametric or fully parametric, and may include baseline covariates that affect the change-time. For this first stage, we exploit existing  methods, theory and efficient  algorithms for interval-censored data  \citep{sun2007statistical,turnbull1976empirical,finkelstein1986proportional, huang1996efficient,anderson2016efficient,wang2016flexible}. In the second stage, we incorporate the first-stage calibration model into the main PH model. We construct a  partial likelihood using the conditional hazard  with respect to the available history, which, in the CRC and aspirin dataset, includes the timing of the last questionnaire, corresponding aspirin-taking status and baseline covariates.

Unlike existing methods, our flexible approach allows to include variables potentially related to the aspirin-taking status, utilizing the available data and subject-matter knowledge in a novel way. In addition, participants without data about post-diagnosis aspirin status, due to e.g., death  or censoring before submitting answers for the first post-diagnosis questionnaire, can be included in the analysis. Existing methods \cite[e.g.,][]{goggins1999applying} exclude these participants from the analysis. 

Our goal in this paper is  three-fold. First, we develop a conceptual framework for the analysis of time-to-event data under the common practice of infrequent updates of a binary covariate of interest. Second, we present a  rigorous analysis of data arisen from several impactful studies in the area of colorectal cancer.  Third, we provide the \textbf{R} package \texttt{ICcalib} that implements our flexible methodology under a wide range of models.  

The rest of the paper is structured as follows. In Section \ref{Sec:DataDesc}, we  describe  the motivating studies. In Section \ref{Sec:MainModel}, we define the main model, and in Section \ref{Sec:Calib},  we present our calibration and risk-set calibration methods. Section \ref{Sec:Asy} contains asymptotic properties where proofs are deferred to the Appendix and the Supplementary Materials. In Section \ref{Sec:Sims}, we describe our  simulation study. In Section \ref{Sec:DataAnalysis}, we present the analysis of the CRC data and in Section \ref{Sec:CMV}, we reanalyze  the CMV data of \cite{goggins1999applying}.  We end with concluding remarks in Section \ref{Sec:Discuss}. 

\section{Data description}
\label{Sec:DataDesc}
 The data were formed from two large cohorts: the NHS, to which 121,701 female nurses enrolled in 1976, and the HPFS that began in 1986 with the enlistment of 51,529 males in various health professions. Description of the studies, the data collection process and eligibility conditions for inclusion in the CRC survival studies can be found in \cite{chan2009aspirin,liao2012aspirin} and \cite{hamada2017aspirin}.   
 
Participants have been receiving questionnaires biennially. During each questionnaire cycle, participants returned their answers in varying times.   CRC researchers are  interested in the first 5 or 10 years following the diagnosis since, conditionally on  10-year survival, the survival rate among CRC patients is similar to the survival rate in the CRC-free population. In our analyses, we considered  10 years follow-up.  About 85\% of stage 4 participants died within the first 5 years (163 of 190), and 108 of them (57\%) had no data on aspirin-taking. Therefore, in order to not bias or destabilize the analysis results,  we limited our analyses to  the  1,371 participants with  stage 1-3 CRC or missing stage data. In the span of 10 years, 249 CRC-related deaths were observed among these participants. Table 1 presents basic descriptive statistics of the main variables. Data on post-diagnosis aspirin use were not available for 113 participants due to loss to follow-up or death  before answering any post-diagnosis questionnaire. Our methodology allowed us to include them in the analysis. 
  \begin{table}[ht]
  	\caption{	\small Summary of the main variables in the dataset \label{Tab:Table1}}
  	\begin{center}
  		 	\small
  		\begin{tabular}{ccccc}
  			\hline
  			& All Data & low CD274 & PIK3CA & PTGS2 \\ 
  			\hline
  			$n$ (No. Events) & 1371 (249) & 278 (50) & 171 (28) & 672 (125) \\ 
  			Age at diagnosis: 	Mean (SD) & 69.4 (9) & 69.4 (8.97) & 69.9 (9) & 68 (8.73) \\ 
  			  			CRC Stage &&&&\\
  			I & 375 (27\%) & 74 (27\%) & 55 (32\%) & 177 (26\%) \\ 
  			II & 453 (33\%) & 99 (36\%) & 62 (36\%) & 218 (32\%) \\ 
  			III & 387 (28\%) & 74 (27\%) & 44 (26\%) & 199 (30\%) \\ 
  			Missing & 156 (11\%) & 31 (11\%) & 10 (6\%) & 78 (12\%) \\ 
  			Pre-diagnosis Aspirin Status &&&&\\
  			Taking & 587 (43\%) & 118 (42\%) & 74 (43\%) & 272 (40\%) \\ 
  			Non-taking & 784 (57\%) & 160 (58\%) & 97 (57\%) & 400 (60\%) \\ 
  			No. Available Questionnaires: 	Mean (SD) & 3 (1.7) & 3.1 (1.66) & 3.2 (1.65) & 3.3 (1.68) \\ 
  			No.  Participants with no Questionnaires & 113 & 21 & 11 & 47 \\ 
  		\end{tabular}
  	\end{center}
  \end{table}

 In the three CRC studies \citep{chan2009aspirin,liao2012aspirin,hamada2017aspirin}, evidence was presented for differential association of aspirin-taking with CRC mortality across the following molecular subtypes, by considering the interaction terms of the aspirin-taking status and these molecular subtypes. The three	 molecular subtypes were  \textit{PTGS2} (cyclooxygenase-2) overexpression \citep{chan2009aspirin},  \textit{PIK3CA} mutation \citep{liao2012aspirin} and low \textit{CD274} (PD-L1)  expression \citep{hamada2017aspirin}. These subtype definitions are not exclusive; a tumor may be included in zero, one, two or three of these subtypes.   The goal of our  analyses  was to evaluate the association between post-diagnosis aspirin  and CRC-related mortality within the subpopulation of each tumor subtype. Therefore, each analysis included different, although overlapping, subgroup of  patients with the subgroup defined by the  molecular subtype for which evidence for aspirin effect was previously revealed. We also included an analysis of all tumors, regardless of the tumor subtype classification. 
 
\begin{figure}
 	\begin{center}
 		\includegraphics[width = 0.65\textwidth]{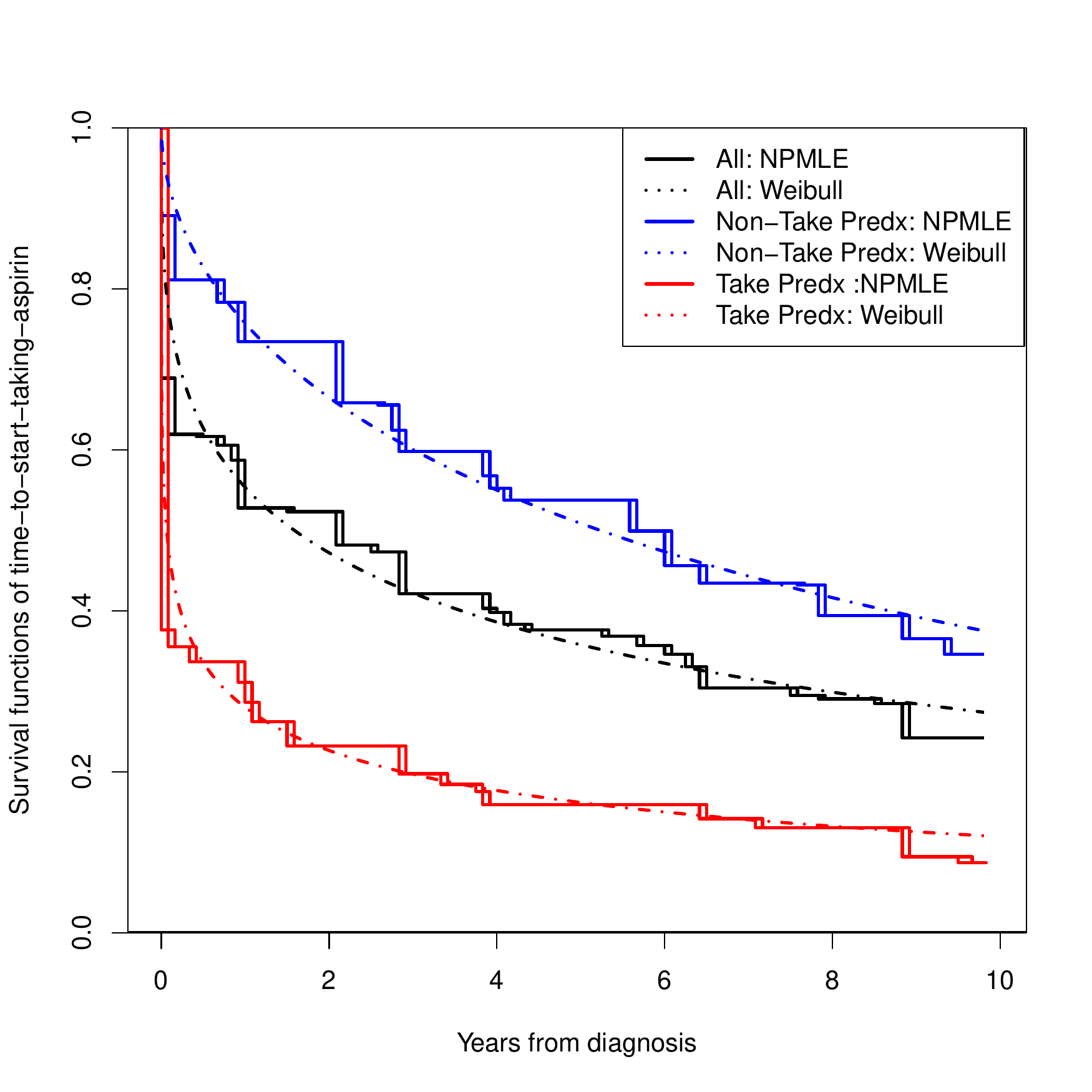}
 	\end{center}
 	\caption{Survival curve estimation of time to start-taking-aspirin for the entire sample (in black, middle curve) and by pre-diagnosis (Predx) aspirin-taking status (takers vs non-takers; bottom red vs top blue curves). Solid lines are the NPMLE, dashed lines are survival curves obtained from Weibull model fitting.}
 	\label{Fig:PrdxAdjAsp}
 \end{figure}

 The post-diagnosis time when diagnosed participants started taking aspirin varied. Figure \ref{Fig:PrdxAdjAsp} presents estimated time-to-start-taking-aspirin curves, for the entire sample (black, middle curve) and by pre-diagnosis aspirin-taking status (red, top curve and blue, bottom curve). The figure depicts  the non-parametric maximum likelihood estimator (NPMLE) and the survival curve estimated assuming Weibull distribution, both calculated from the interval-censored data about post-diagnosis aspirin-taking \citep{sun2007statistical}. With the exception of emergency surgeries, patients diagnosed with CRC were asked to stop taking aspirin to prepare to the surgery, and were allowed to resume  aspirin few days after the surgery.    It is likely that patients who had been taking aspirin prior to  CRC diagnosis were more inclined  to start taking aspirin, once it was possible.  This is consistent with the rapid drop right after the baseline in the curve  for pre-diagnosis aspirin-takers  in Figure \ref{Fig:PrdxAdjAsp}.
\section{The main model}
\label{Sec:MainModel}
Let $T$  be the time to event of interest, in the CRC data time-to-death following CRC diagnosis.   $T$ is possibly right censored by a censoring time $C$. We assume that, conditionally on measured baseline covariates, $T$ and $C$ are independent, and  let $\tilde{T}=\min(T,C)$ and $\delta= I\{T<C\}$ be the observed time and  censoring indicator, respectively. Of main interest is the association of the monotone nondecreasing binary covariate $X(t)$ with $T$. For presentational simplicity, we henceforth refer to $X(t)$ as the \textit{exposure}, although it can be a treatment or any other binary covariate. Let $V$ be the change time in the value of $X$. That is, $V$ is the first time when $X(t)=1$. As described in the introduction, $X(0)=0$ and  $X(t)=1$ for all $t\ge V$.  In some applications, the exposure may change its value back to zero, at some point after $V$.  In studying aspirin-taking this option  is often neglected. This is done to avoid misinterpretation of the results due to reverse causality, because deteriorating patients, who are more likely to die, may stop taking aspirin. In our study of aspirin-taking and survival among CRC patients, we can therefore refer to the exposure as ever aspirin-taking (since diagnosis). 

 Let $\bZ$ be a column vector of covariates presumably associated with $T$.   For simplicity of presentation, we will assume that $\bZ$ is time-independent. The proposed methods can be straightforwardly applied to the scenarios when $\bZ$ is time-dependent. The  Cox PH model \citep{cox1972regression} for the hazard function of $T$ given $X(t)$ and $\bZ$ is 
  \begin{equation}
  \label{Eq:PHModel}
  \lambda(t|X,\bZ)=\lim\limits_{\Delta\rightarrow0}\Delta^{-1}P(t \le T < t + \Delta|X(t), \bZ, T \ge t	)=\lambda_0(t)\exp(\beta X(t)+ \bgamma^T\bZ),
  \end{equation} 
where $\lambda_0(t)$ is an unspecified baseline function and $\beta$  and $\bgamma$ are parameters to be estimated. We are mainly interested in $\beta$.  We  refer to this PH model for $\lambda(t|X,\bZ)$ as the \textit{main model}. We assume that, given $X_i(t)$, $\{X_i(s), s< t\}$ are not informative about the hazard at time $t$.  The partial likelihood for $(\beta,\bgamma)$ is 
\begin{equation}
\label{Eq:PartLikStand}
L(\beta,\bgamma)=\prod_{i=1}^{n}\left[\frac{\exp(\beta X_i(t_i)+ \bgamma^T\bZ_i)}{\sum\limits_{j=1}^{n}Y_j(t_i)\exp(\beta X_j(t_i)+ \bgamma^T\bZ_j)}\right]^{\delta_i}
\end{equation} 
 where $Y_i(t)=I\{\tilde{T}_i\ge t\}$ is an at-risk indicator. 
 
If $V_i$ was known for all $i=1,...,n$,  we could simply set the time-dependent exposure to be equal to zero for all $t<V$ and to one after that time, i.e.,  $X_i(t)=I\{t\ge V_i\}$.  However, the data collected from questionnaires are often limited.  The aspirin-taking status $X_i$ was measured only at a series of $M_i$ discrete time points:  $0=w_{i0} < w_{i1} <...< w_{iM_i}$, with $X(w_{i0})=0$. If $X_i$ was not measured at any time point before $\tilde{T}_i$, then $M_i=0$. In the CRC data, these time points are the times when questionnaires were answered by the participants. When the main event is terminal, as in our data, questionnaire data available only until the event or censoring time; that is,  $w_{iM_i}\le\tilde{T}_i$. In other studies, e.g., the CMV data previously analyzed by \cite{goggins1999applying},  data on the exposure can be obtained even after the main event has occurred.  The time points at which $X$ was measured are either fixed or random. We take a working assumption that their distribution is independent of the other random variables presented so far. This assumption would not hold in certain situations. The degree of the deviation from this assumption, and its impact on the proposed methodology, will be discussed later in the paper.

Let $\overline{w}_i^t$ denote the last time $X_i$ was measured before time $t$; $\overline{w}_i^t=0$ if $X_i$ was not previously measured. If $X(\overline{w}_i^t)=1$, then $X_i(t)=1$.  However, if $X(\overline{w}_i^t)=0$, then $X_i(t)$ can be either one or zero.  Therefore, the likelihood \eqref{Eq:PartLikStand} cannot be calculated from the observed data.  Two common methods for addressing this type of missing data are the last-value-carried-forward (LVCF) and midpoint imputation (MidI) methods.  

The  LVCF  method imputes missing values as the last observed values. In our context, this is equivalent to setting  $X_i(t)=X_i(\overline{w}_i^t)$. That is, a  participant is assumed to not taking aspirin, until the first time she reports aspirin-taking.  The MidI method imputes the changepoint (i.e., $V$) in the middle of the interval in which the changepoint is known to have  occurred. For example, if a participant reported she was not taking aspirin 1 year after diagnosis, and then reported taking aspirin 2.5 years after diagnosis, then MidI would assume  $V=1.75$ for that participant; LVCF would assume $V=2.5$.  These are ad hoc methods that may lead to substantially biased results. This calls for new conceptual framework and methodology development, which we now present.

\section{A calibration approach}
\label{Sec:Calib}
 Let	$\mathcal{F}_{it}=\{\bZ_i, \bQ_i, \overline{w}^t_{i}, X_i(\overline{w}^t_{i}), \{T_i\ge t\}\}$ be the  history of an at-risk participant $i$ until time $t$, where $\bQ_i$ is a possibly vector-valued covariates informative about $X_i(t)$,   and $\{T_i\ge t\}$ reflects the fact that participant $i$ has been event-free so far. In our application, $\bQ_i$ includes, among other covariates, the pre-diagnosis aspirin-use status, which is clearly informative about $X_i(t)$, as previously demonstrated in Figure \ref{Fig:PrdxAdjAsp}.   If a covariate affects both $T_i$ and $X_i$, it is included in both $\bZ_i$ and $\bQ_i$. 
 
 Under model \eqref{Eq:PHModel},   the hazard function conditional on $\mathcal{F}_{it}$ is
\begin{align*}
\lambda_i(t|\mathcal{F}_{it})&=\lim\limits_{\Delta\rightarrow 0}\Delta^{-1}P(t \le T_i < t + \Delta|\mathcal{F}_{it})\\
&=\lim\limits_{\Delta\rightarrow 0}\Delta^{-1}E_{X_i(t)|\mathcal{F}_{it}}[P(t \le T_i < t + \Delta|T_i\ge t, X_i(t), \bZ_i)]\\
&=\lambda_0(t)\exp(\bgamma^T\bZ_i)E\bigl[\exp(\beta X_i(t))|\mathcal{F}_{it}\bigr],
\end{align*}
following the assumption that conditionally on $X_i(t),\bZ_i$ and $\{T_i\ge t\}$, the probability of event at time $t$ is independent of $\bQ_i$,  $\overline{w}^t_{i}$ , and  $X_i(\overline{w}^t_{i})$. 

 A similar derivation was presented  for the case of measurement error in a time-dependent continuous covariate  \citep{prentice1982covariate}, and is exploited by the ``regression calibration'' method \citep{wang1997regression,ye2008semiparametric,liao2011survival}. It is readily seen that  $\lambda_i(t|\mathcal{F}_{it})$ is also a PH model \citep{prentice1982covariate} and therefore we can consider the partial likelihood
\begin{equation}
\label{Eq:PartLikF}
L^{\mathcal{F}}(\beta,\bgamma)=\prod_{i=1}^{n}\left[\frac{\exp(\bgamma^T\bZ_i)E\bigl[\exp(\beta X_i(t_i))|\mathcal{F}_{it_i}\bigr]}{\sum\limits_{j=1}^{n}Y_j(t_i)\exp( \bgamma^T\bZ_j)E\bigl[\exp(\beta X_j(t_i))|\mathcal{F}_{jt_i}\bigr]}\right]^{\delta_i}.
\end{equation}
%In measurement error models, and in joint-modeling of time-to-event and longitudinal measurements, similar partial likelihoods  have arisen. Often, the next step involves an approximation of  the expectations in  \eqref{Eq:PartLikF}, e.g.,  by inserting the expectation into the power of the exponent, and then replacing the expectation of the exposure given some observed relevant data, with a prediction from a relevant model. \cite{prentice1982covariate} has shown that this approach is a form of ``regression calibration''. \cite{wang1997regression} described the general methodology for applying regression calibration in time-to-event data. A semiparametric mixed-model calibration was suggested by \cite{ye2008semiparametric}.   
Unlike the regression calibration method,  the fact that $X_i(t)$ is binary allows us to explicitly write the expectation in \eqref{Eq:PartLikF}   as
\begin{equation}
\label{Eq:MGF}
	E\bigl[\exp(\beta X_i(t))|\mathcal{F}_{it}\bigr] = 1 +	P(X_i(t)=1|\mathcal{F}_{it})(\exp(\beta)-1),
\end{equation}
where $P(X_i(t)=1|\mathcal{F}_{it})$ can be expressed using the distribution of $V$, the time of exposure status change,
\begin{equation}
\label{Eq:PrV}
P(X_i(t)=1|\mathcal{F}_{it})= \left\{ 
\begin{array}{cc} 
1 &  X_i(\overline{w}^{t}_{i})=1\\ 
\dfrac{P(\overline{w}^{t}_{i} < V_i  \le t |\bQ_i, T_i\ge t)}{P(\overline{w}^{t}_{i} < V_i|\bQ_i, T_i\ge t)} & X_i(\overline{w}^{t}_{i})=0  
\end{array}\right..
\end{equation}
In words, the probability of a positive aspirin-taking  status at time $t$, conditionally on the personal history, equals to one, if the participant previously reported   she is taking aspirin, and if she did not report aspirin-taking previously, it  equals to the probability of a change in the aspirin-taking status between the last questionnaire time  and time $t$, conditionally on no aspirin-taking at time of last questionnaire. Combining \eqref{Eq:MGF} and \eqref{Eq:PrV}, it is evident that $E\bigl[\exp(\beta X_i(t))|\mathcal{F}_{it}\bigr]$ is a functional of the distribution of $V$ conditionally on $\bQ_i$ and $\{T_i\ge t\}$.  If the distribution of $V_i|\bQ_i,\{T_i\ge t\}$ was known, we could have obtained valid estimates for $\beta$ and $\bgamma$ by substituting  \eqref{Eq:MGF} and \eqref{Eq:PrV}  into $L^{\mathcal{F}}$.  However, the distribution of $V_i|\bQ_i,\{T_i\ge t\}$ is unknown.

Let  $\mathcal{G}_{it}$ be $\mathcal{F}_{it}$ without the survival information, i.e., $\mathcal{G}_{it}=\{\bZ_i, \bQ_i, \overline{w}^t_{i}, X_i(\overline{w}^t_{i})\}$. Our first proposal is to estimate $\beta$ and $\bgamma$ by applying two modifications to $L^{\mathcal{F}}$. First, we  replace $\mathcal{F}$ by $\mathcal{G}$ in the expectations in \eqref{Eq:PartLikF} to get   $E\bigl[\exp(\beta X_i(t))|\mathcal{G}_{it}\bigr]$.  The second modification involves replacing expectations of the form $E\bigl[\exp(\beta X_i(t))|\mathcal{G}_{it}\bigr]$  by  estimators  $\widehat{E}\bigl[\exp(\beta X_i(t))|\mathcal{G}_{it}\bigr]$.  Our ordinary calibration (OC) estimator $\hat{\beta}_{OC}$ is the maximizer of 
  \begin{equation}
  \label{Eq:PartLikG}
  L^{\mathcal{G}}(\beta,\bgamma)=\prod_{i=1}^{n}\left[\frac{\exp(\bgamma^T\bZ_i)\widehat{E}\bigl[\exp(\beta X_i(t_i))|\mathcal{G}_{it_i}\bigr]}{\sum\limits_{j=1}^{n}Y_j(t_i)\exp( \bgamma^T\bZ_j)\widehat{E}\bigl[\exp(\beta X_j(t_i))|\mathcal{G}_{jt_i}\bigr]}\right]^{\delta_i},
  \end{equation}
  with respect to $\beta$ and $\bgamma$. Noting the simplicity of this likelihood function, maximization can be done in a  straightforward way,  e.g., using Newton-Raphson algorithm. 
  
 The expectation $E\bigl[\exp(\beta X(t))|\mathcal{G}_{t}\bigr]$ can be expressed as a functional of the distribution of $V|\bQ$, as in  \eqref{Eq:MGF} and \eqref{Eq:PrV}, with $\mathcal{F}$ replaced by $\mathcal{G}$, and omitting $\{T_i\ge t\}$ in \eqref{Eq:PrV}. 
 Therefore, we first estimate  the distribution of $V|\bQ$, and then calculate  $\widehat{E}\bigl[\exp(\beta X_i(t))|\mathcal{G}_{it}\bigr]$  using the estimated distribution.  
 \subsection{Calibration models fitted from interval-censored data}
 \label{SubSec:OC}
 For each participant, the data available on $V$ include the previously mentioned measurement times $0=w_{i0} < w_{i1} <...< w_{iM_i}$, and the corresponding measurements. These data can be summarized in a form of the interval $V$ is censored into,  denoted by $(w_{iL},w_{iR}]$, where 
\begin{align*}
w_{iL}&=\left\{\begin{array}{cc}
0 & M_i=0 \quad \text{or} \quad  X(w_{i1})=1\\
\max_j\{w_{ij}:X(w_{ij})=0\} & M_i>0 \quad \& \quad X(w_{i1})=0 \\
\end{array}\right. ,\\
 w_{iR}&=\left\{\begin{array}{cc}
\min_j\{w_{ij}:X(w_{ij})=1\} & M_i>0 \quad \& \quad  X(w_{iM_i})=1\\
\infty & M_i=0 \quad \text{or} \quad X(w_{iM_i})=0 \\
\end{array}\right. .
\end{align*}
If $w_{iL}=0$,  $V_i$ is left-censored, and if $w_{iR}=\infty$, $V_i$ is right-censored. Note that data about measurements of $X_i$ before $w_{iL}$ or after  $w_{iR}$ do not add information about $V_i$.

We  refer to the model for the distribution of $V$ as the \textit{calibration model}. Let $S^V(v)=P(V>v)$ be the survival function of $V$. The general form of the likelihood of interval-censored time-to-event data, under independent censoring assumption, is \citep{sun2007statistical} 
\begin{equation}
\label{Eq:IClik}
L^V=\prod_{i=1}^{n}P(w_{iL} < V_i \le w_{iR})=\prod_{i=1}^{n}[S^V(w_{iL})-S^V(w_{iR})].
\end{equation}
The NPMLE estimator has been studied as an extension of the Kaplan-Meier estimator to interval-censored data.  Algorithms suggested to find the NPMLE include the self-consistency EM algorithm \citep{turnbull1976empirical}, the iterative convex minorant (ICM) algorithm \citep{groeneboom1992information}, and an EM-ICM hybrid algorithm  \citep{wellner1997hybrid}. Consistency and asymptotic distribution (at a $n^{1/3}$ rate) were proved by \cite{groeneboom1992information}. 

One can  postulate a parametric or semiparametric model for the distribution of $V$. This is especially appealing when the distribution of $V$ is likely to depend on additional covariates, previously denoted by $\bQ$. In the CRC studies,  pre-diagnosis aspirin-taking status is strongly associated with the time from CRC diagnosis to post-diagnosis aspirin-taking (Figure \ref{Fig:PrdxAdjAsp}).  Additional covariates   include risk factors for cardiovascular and cerebrovascular events because aspirin is often taken to reduce the risk of these events among high-risk patients; see Section \ref{Sec:DataAnalysis} for further discussion.   Therefore, while our framework allows for variety of time-to-event models, we focus in this paper on a calibration model with covariates, and specifically a PH model of this nature. 

 Let $\beeta$ denote a vector of parameters characterizing the distribution of $V$. An estimator of  $\beeta$ is obtained by maximizing  the equivalent of  \eqref{Eq:IClik} under a model for $S^V$ that possibly accommodates the  covariates. See Chapter 2 of \cite{sun2007statistical} for discussion of parametric models for interval-censored time-to-event data. A more flexible model is a PH regression model with an unspecified baseline hazard function. \cite{finkelstein1986proportional} discussed PH models for interval-censored data and suggested discretization of the baseline hazard function. Algorithms for computation of the MLE were previously proposed \citep{huang1996efficient,huang1997interval,pan1999extending}, and asymptotic theory was studied  \citep{huang1996efficient,huang1997interval}.  \cite{cai2003hazard} developed a related method for fitting PH models from interval-censored data  using a linear spline model for the log-baseline hazard and maximized a penalized log-likelihood.

 We adopt the recently developed framework  of \cite{wang2016flexible}, which uses flexible I-splines \citep{ramsay1988monotone} for the cumulative baseline hazard function. \cite{wang2016flexible} further developed a fast EM algorithm, which we apply in our simulations and data analysis. Let $\Lambda^V_0(v)$ and $S^V_{\beeta, \Lambda^V_0}(v|\bQ)$  be the cumulative baseline hazard function and the survival function of $V$, respectively, which under the PH model are 
 \begin{align}
 \begin{split}
 \label{Eq:PHIsplines}
 S^V_{\beeta,\Lambda^V_0}(v|\bQ)&=\exp[-\Lambda^V_0(v)\exp({\bpsi}^T\bQ)],\\
 \Lambda^V_0(v)&=\sum\limits_{k=1}^{K}\alpha_kb_k(v),
  \end{split}
 \end{align} 
 where for all $k=1,...,K$,  $\alpha_k\ge0$ are unknown parameters to be estimated and  $b_k(v)\in[0,1]$ are  integrated spline basis functions, that are nondecreasing. The spline basis functions are calculated  according to the  user specification of  $m$ interior  knots and  a polynomial degree for the  basis functions.  The resulting $\Lambda^V_0(v)$ is guaranteed to be monotone increasing.  See \cite{ramsay1988monotone} and \cite{wang2016flexible} for further details and discussion about I-splines in general and for the PH model, respectively, including the important practical issues of choosing the number and the location of the knots. 
 
  To summarize this section, the calibration model is fitted by maximizing the likelihood \eqref{Eq:IClik} while substituting a model for $S^V$, chosen based on the available data and subject-matter knowledge. In our case, we are going to use the PH model \eqref{Eq:PHIsplines} for the time until aspirin-taking following a CRC diagnosis. 
   \subsection{Risk-set calibration}
The OC estimator $\hat{\beta}_{OC}$ may suffer from asymptotic bias. It is calculated as the maximizer of $L^\mathcal{G}$ while  the partial likelihood is $L^\mathcal{F}$. The degree of  divergence between  $L^\mathcal{F}$ and $L^\mathcal{G}$ depends on how different $E\bigl[\exp(\beta X(t))|\mathcal{F}_{t}\bigr]$ and  $E\bigl[\exp(\beta X(t))|\mathcal{G}_{t}\bigr]$ are. Recall that $\mathcal{G}_{t}$ was defined by omitting from $\mathcal{F}_{t}$ the event $\{T\ge t\}$. If the probability of this event is close to one, as in the case of rare events, the bias should be attenuated. If $X$ has no effect on $T$, i.e.,  under the null,  $X(t)$ is independent of $\{T\ge t\}$ and $E\bigl[\exp(\beta X(t))|\mathcal{F}_{t}\bigr]=E\bigl[\exp(\beta X(t))|\mathcal{G}_{t}\bigr]$. This implies that  as the absolute value of $\beta^0$, the true value of $\beta$, increases, a larger bias may be expected. 

Another source of bias stems from fitting the calibration model under the independent interval-censoring assumption,  which is the assumption that $w_{iL}$ and $w_{iR}$ are independent of $V_i$. However, in our studies, the time to event, $T$, is informative about the censoring in the calibration model.  If, for example, $X$ reduces the risk of death, the time of aspirin-taking status change is more likely to be censored in non-aspirin-taking patients. This may  cause bias in the estimation of $\beeta$ when fitting the calibration model. As before, if the event is rare,   the censoring of $V$ is most likely due to administrative reasons, and hence the independent interval-censoring assumption approximately holds. Furthermore, under the null, the censoring interval is independent of $V$. As before, larger $|\beta^0|$ typically implies more substantial bias. We investigate this point in the simulation studies in Section \ref{Sec:Sims} and the Supplementary Materials. In studies with non-terminal  event, the independent censoring assumption for the calibration model may be more plausible, because data on $X$ can be collected after the main event has occurred.

In order to reduce potential bias, we propose a risk-set calibration (RSC) procedure,  an adaption of risk-set regression calibration previously developed in the context of error-prone covariates in survival analysis \citep{xie2001risk,ye2008semiparametric,liao2011survival}. This method uses $L^\mathcal{F}$, and estimate the distribution of  $V|\bQ, T\ge t$ by refitting the calibration model for $V|\bQ$ at each observed event time, using only the members of the risk set at that time, so only participants with $T\ge t$ are used. Then, at each risk set, we plug-in the  estimated distribution of $V|\bQ, T\ge t$ in \eqref{Eq:PrV}, leading to $\widehat{P}(X(t)|\mathcal{F}_{t})$ which is then  substituted in \eqref{Eq:MGF} to obtain $\widehat{E}\bigl[\exp(\beta X(t))|\mathcal{F}_{t}\bigr]$ for $L^\mathcal{F}$.

The RSC is expected to lead to less bias than OC, especially when $|\beta^0|$ is large \citep{xie2001risk}.  However,  some asymptotic bias in the RSC estimator may be expected, due to model misspecification. Even if the PH model for the distribution of $V|Q, T\ge t$ holds at $t=0$, it is not expected to hold for all $t>0$. The RSC estimator is also expected to have larger variance, due to increased number of parameters, and the decreasing (in $t$) sample size for the risk-set calibration models.  Therefore, it is advised to use this estimator when the $X$-$T$ association is strong, and the sample size is large. 
\section{Asymptotic properties}
\label{Sec:Asy}
We focus on the PH model under the I-splines representation  for $\Lambda^V_0$ of \cite{wang2016flexible}. The results can be straightforwardly extended to parametric models for $S^V$. Let $\btheta=(\beta, \bgamma)$ and $\beeta=(\bpsi, \balpha)$ and recall that  $\hat{\beta}_{OC}$ is obtained by maximizing $L^{\mathcal{G}}$, or alternatively, by solving $U^{\mathcal{G}}(\btheta,\hat{\beeta})=0$ where 
\begin{equation*}
\label{Eq:U}
\bU^{\mathcal{G}}(\btheta;\beeta)=\frac{1}{n}\sum\limits_{i=1}^{n}\int\limits_{0}^{\tau}\left[\ba_i(\btheta,\beeta,t)-\frac{\bS^{(1)}(\btheta,\beeta,t)}{S^{(0)}(\btheta,\beeta,t)}\right]dN_i(t),
\end{equation*}
with $\tau$ being the study end-time, $N_i(t)$  the counting process associated with $T_i$, and $\ba_i, S^{(0)}$ and $\bS^{(1)}$ are as defined in the Appendix.   Under assumptions \ref{Ass:knots}--\ref{Ass:Hess} in the Appendix, $\hat{\btheta}_{OC}\xrightarrow{P}\btheta^\star$. The proof is outlined in the Appendix.  Furthermore,
$
\sqrt{n}(\hat{\btheta}-\btheta^\star)\xrightarrow{D}N(0,\mathcal{V})
$
and $\mathcal{V}$ can be estimated by a sandwich estimator
\begin{equation}
\label{Eq:VarEst}
\hat{\mathcal{V}}= [-\nabla_{\btheta}\bU^{\mathcal{G}}_{\btheta}(\hat{\btheta}_{OC},\hat{\beeta})]^{-1}\left(\frac{1}{n}\sum\limits_{i=1}^{n}\hat{\br}_i(\hat{\btheta}_{OC},\hat{\beeta})\hat{\br}_i^T(\hat{\btheta}_{OC},\hat{\beeta})\right)[-\nabla_{\btheta}\bU^{\mathcal{G}}_{\btheta}(\hat{\btheta}_{OC},\hat{\beeta})]^{-1}
\end{equation}
with $\hat{\br}_i$ being a sample version of $\br_i$ given in the Appendix. The proof is outlined in the Appendix.

For the RSC estimator, results of similar nature are given and proved in the Supplementary Materials. Main modifications in the results are that $\hat{\btheta}_{RSC}\xrightarrow{P}\btheta^{\star\star}$, where $\btheta^{\star\star}$ is possibly different from $\btheta^\star$, and that $\beeta$ is replaced by $\tilde{\beeta}(t)$,  a time-dependent parameter vector.

\section{Simulation study}
\label{Sec:Sims}
We considered a PH calibration model in our main simulation study. In the Supplementary Materials, we describe simulation studies for parametric and nonparametric  calibration models  in  absence of covariates for calibration. 

For the main model, the hazard function of $T$ was $\lambda_0(t)\exp(\beta X(t) + \gamma_1Q_1 + \gamma_2Q_2 + \gamma_3Z_3)$, with $Q_1\sim Bernoulli(0.5), Q_2\sim N(0,0.5^2)$ and $Z_3\sim N(0,1)$, all independent of each other.  We took the Gompertz baseline hazard function $\lambda_0(t)=0.1\exp(0.25t)$,  and $\gamma_1=\log(0.75)$, $\gamma_2=\log(2.5)$ and $\gamma_3=\log(1.5)$. For $\beta$, we considered the values $\beta=\log(1), \log(2), \log(5), \log(7)$. We simulated time-to-event data with time-dependent covariate as described in \cite{austin2012generating}. We took exponential censoring (mean=5) and additional censoring at time 5. The resulting censoring rate varied between 42\% and 63\%, depending on the value of $\beta$.

For the calibration model, we used the setup of \cite{wang2016flexible}, with $S^V(v)=\exp[(\log(1+v)+\sqrt{v})\exp(\eta_1Q_1+ \eta_2Q_2)]$, where $\eta_1=\log(2)$, $\eta_2=\log(0.5)$,  and $Q_1$ and $Q_2$ are the same covariates as in the main model. For each observation, $M^\star$ questionnaire time points were simulated from the intervals on the equally spaced grid of $[0,5]$. For example, under $M^\star=2$, the first questionnaire time point was simulated from $U(0,2.5)$, and the second from $U(2.5,5)$. However, to mimic the motivating studies, we considered a terminal main event, and kept only questionnaire time points before $\tilde{T}_i$.

 The PH calibration model fitting was done for each simulated dataset with $m=5$ equally-spaced interior knots and a quadratic order for the basis functions of the I-spline cumulative baseline hazard.  In the CRC  data analysis (Section \ref{Sec:DataAnalysis}), we have used the data as recommended in \cite{wang2016flexible} to choose more carefully the number of knots, and used cubic splines. Standard errors were estimated by \eqref{Eq:VarEst} and confidence intervals were calculated using the normal distribution. 
   
We simulated 1000 datasets per scenario. Table \ref{Tab:SimRes} summarizes the results  for the LVCF, the PH calibration model (PH-OC) and PH risk-set calibration models (PH-RSC), for $M^\star=2,5$. Under the null, all three methods were valid. As the association between $X$ and $T$ got stronger, a more substantial bias was observed. The OC estimator preformed generally well, with increased bias and lower coverage rates of the 95\% confidence interval for the combination of $\beta^0>\log(2)$ and $M^\star=2$. The RSC estimator had lower bias in these scenarios. These observed biases were much smaller (in absolute value) than the bias of the LVCF estimator. 

Table A.1 in the  Supplementary Materials presents more results from this simulation scenario. It includes the results for $M^\star_i=10$,  $\beta=\log(1/2), \log(1/5), \log(1/7)$, and  the MidI estimator. In terminal main event scenarios,  MidI had a very large bias, even under the null. This is because a finite interval is only observed for observations with left- or interval-censored exposure times. Right-censored exposure times are dealt with differently (e.g., using LVCF) from left- or interval-censored observations, and a right-censored exposure time is the result of the main event occurring before a change in $X$ was observed. This creates negative dependency between the imputed $X$ and $\tilde{T}$, even under the null. In studies with non-terminal events,  the MidI method is valid under the null, but not when $\beta^0\ne 0$. A small simulation study (not presented here) confirmed these claims.
\begin{table}
	\caption{	\footnotesize Simulation study results under a calibration PH model. Methods compared are LVCF, PH calibration model (PH-OC) and PH risk-set calibration models (PH-RSC). We present mean estimates (Mean), empirical standard deviations (EMP.SE), mean estimated standard errors ($\widehat{SE}$) and empirical coverage rate of 95\% confidence intervals (CP95\%) for $\beta$.  Results are based on 1000 simulated datasets per scenario. \label{Tab:SimRes}}
	\begin{center}
			\footnotesize
	\begin{tabular}{ccccccc}
		\hline
		$\beta^0 [\exp(\beta^0)]$  & $M^\star$ & Method & Mean & EMP.SE & $\widehat{SE}$ & CP95\% \\ 
		\hline
		0.000  & 2 & LVCF & -0.002 & 0.141 & 0.135 & 0.944 \\ 
	%	 $[1.00]$ &  & MidI & -0.554 & 0.130 & 0.131 & 0.011 \\ 
		$[1.00]$ &  & PH-OC & 0.003 & 0.183 & 0.178 & 0.950 \\ 
		 &  & PH-RSC & 0.004 & 0.183 & 0.177 & 0.952 \\ 
		 & 5 & LVCF & 0.003 & 0.122 & 0.124 & 0.955 \\ 
%		 &  & MidI & -0.300 & 0.113 & 0.121 & 0.270 \\ 
		 &  & PH-OC & 0.007 & 0.138 & 0.146 & 0.956 \\ 
		 &  & PH-RSC & 0.007 & 0.138 & 0.142 & 0.956 \\ 
		 \hline
		0.693 & 2 & LVCF & 0.462 & 0.119 & 0.118 & 0.498 \\ 
	%	 $[2.00]$ &  & MidI & -0.097 & 0.109 & 0.115 & 0.000 \\ 
		$[2.00]$ &  & PH-OC & 0.680 & 0.175 & 0.179 & 0.936 \\ 
		 &  & PH-RSC & 0.684 & 0.177 & 0.175 & 0.938 \\ 
		 & 5 & LVCF & 0.572 & 0.107 & 0.110 & 0.810 \\ 
%		 &  & MidI & 0.228 & 0.100 & 0.107 & 0.004 \\ 
		 &  & PH-OC & 0.690 & 0.132 & 0.145 & 0.958 \\ 
		 &  & PH-RSC & 0.689 & 0.132 & 0.137 & 0.957 \\ 
		 \hline
		1.609 & 2 & LVCF & 0.968 & 0.110 & 0.109 & 0.000 \\ 
%		 $[5.00]$ &  & MidI & 0.368 & 0.096 & 0.107 & 0.000 \\ 
	$[5.00]$	 &  & PH-OC & 1.472 & 0.179 & 0.210 & 0.869 \\ 
		 &  & PH-RSC & 1.516 & 0.190 & 0.195 & 0.897 \\ 
		 & 5 & LVCF & 1.212 & 0.096 & 0.099 & 0.013 \\ 
%		 &  & MidI & 0.787 & 0.086 & 0.097 & 0.000 \\ 
		 &  & PH-OC & 1.577 & 0.139 & 0.166 & 0.951 \\ 
		 &  & PH-RSC & 1.575 & 0.137 & 0.151 & 0.948 \\ 
		 \hline
		1.946 & 2 & LVCF & 1.130 & 0.112 & 0.110 & 0.000 \\ 
%		 $[7.00]$ &  & MidI & 0.501 & 0.094 & 0.108 & 0.000 \\ 
	 $[7.00]$	 &  & PH-OC & 1.695 & 0.182 & 0.230 & 0.723 \\ 
		 &  & PH-RSC & 1.773 & 0.198 & 0.205 & 0.816 \\ 
		 & 5 & LVCF & 1.410 & 0.097 & 0.098 & 0.000 \\ 
	%	 &  & MidI & 0.939 & 0.086 & 0.096 & 0.000 \\ 
		 &  & PH-OC & 1.890 & 0.148 & 0.187 & 0.933 \\ 
		 &  & PH-RSC & 1.890 & 0.147 & 0.158 & 0.929 \\ 
		\hline
	\end{tabular}
	\end{center}
\end{table}

To investigate the performance of our methods  in other settings, we have considered additional simulation studies without any covariates affecting the time-to-exposure (i.e., no $Q_1$ and $Q_2$ in the calibration model) and compared Weibull and nonparamteric calibration models when the true calibration model was Weibull, and when it was piecewise Exponential.  The results, presented in the Supplementary Materials,  generally agreed with the results we have descried in this section.
\section{Results of aspirin and CRC survival analyses}
\label{Sec:DataAnalysis}

Our first step was to construct a calibration model. That is, a model for the time-to-start-aspirin-taking. The 113 participants without  available questionnaire data were not used for fitting the calibration model, and the calibration model was fitted using the remaining 1,258 participants. The baseline potential covariates included gender, age-at-diagnosis, pre-diagnosis body mass index (BMI), pre-diagnosis aspirin-taking status,  and the following tumor  characteristics:  disease stage (1-3 and missing), differentiation (poor vs well-moderate) and  location (proximal colon, distal colon or rectum). We considered three PH calibration models: (I) a model with all the aforementioned covariates, (II) a model with all non tumor-related baseline covariates and disease stage, (III) a model with all non tumor-related baseline covariates.   Model (III) minimized the BIC \citep{schwarz1978estimating}. In addition, including strong determinant of the terminal event, such as disease stage, is  undesired. An association between disease stage and aspirin-taking could be the result of violation of the independent censoring assumption  for the calibration model. That is, the association of the disease stage with aspirin-taking could be only due to the  association of disease stage with the censoring event, i.e., death. 

% and then the tumor stage is associated with time-to-start-taking-aspirin through its association with the censoring event, death. 
Model (III) is  logically sound from a subject-matter perspective. Aspirin is a preventive care for patients in high-risk for vascular diseases. Therefore, determinants of  vascular diseases would also increase the probability of aspirin-taking. The covariates in Model (III), age, gender, and BMI, are all well-established risk factors for vascular diseases. Therefore, we adopted Model (III) as our calibration model. Table \ref{Tab:CalibModel} presents the results of fitting this PH model to the data. We used the same PH calibration  model  for all the analyses since there is no reason to believe that the tumor subtype, that was not even known at the time of diagnosis, informs the  aspirin-taking status changing time.

As suggested by \cite{wang2016flexible}, we used BIC to choose the number of equally-spaced interior knots, which led to $m=11$. We used the flexible cubic order for the spline basis functions. The number of interior knots according to the AIC criterion \citep{akaike1974new} was $m=25$.  The final results did not substantially change when we specified $m=25$ instead of $m=11$ interior knots. 
 \begin{table}
 	\caption{The PH calibration model for time-to-start-taking-aspirin, using cubic order and $m=11$ interior knots. Covariates include  pre-diagnosis aspirin status (Predx-asp, taker versus non-taker), pre-diagnosis BMI (Predx-BMI), age-at-diagnosis (Age-at-dx) and  gender (Female). \label{Tab:CalibModel}}
\begin{center}
 	\small
 	\begin{tabular}{ccccc}
 		\hline
 		& Est ($\widehat{SE}$) & $\widehat{HR}$ & CI & $p$-value \\ 
 		\hline
 		Predx-asp & 1.14 (0.074) & 3.11  & $[2.69, 3.60]$ &  $<0.001$ \\ 
 		Predx-BMI & 0.03 (0.006) & 1.03  & $[1.02, 1.04]$ &  $<0.001$ \\ 
 		Age-at-dx & 0.01 (0.001) & 1.01  &  $[1.007, 1.013]$ & $<0.001$ \\ 
 		Female & -0.18 (0.073)  & 0.83 &    $[0.72, 0.96]$ & 0.013 \\ 
 		\hline
 	\end{tabular}
 	\end{center}
 \end{table}
 
\begin{figure}[t]
\begin{center}
	\captionsetup[subfigure]{justification=centering}
	\begin{subfigure}[t]{0.45\textwidth}
				\includegraphics[width=\textwidth]{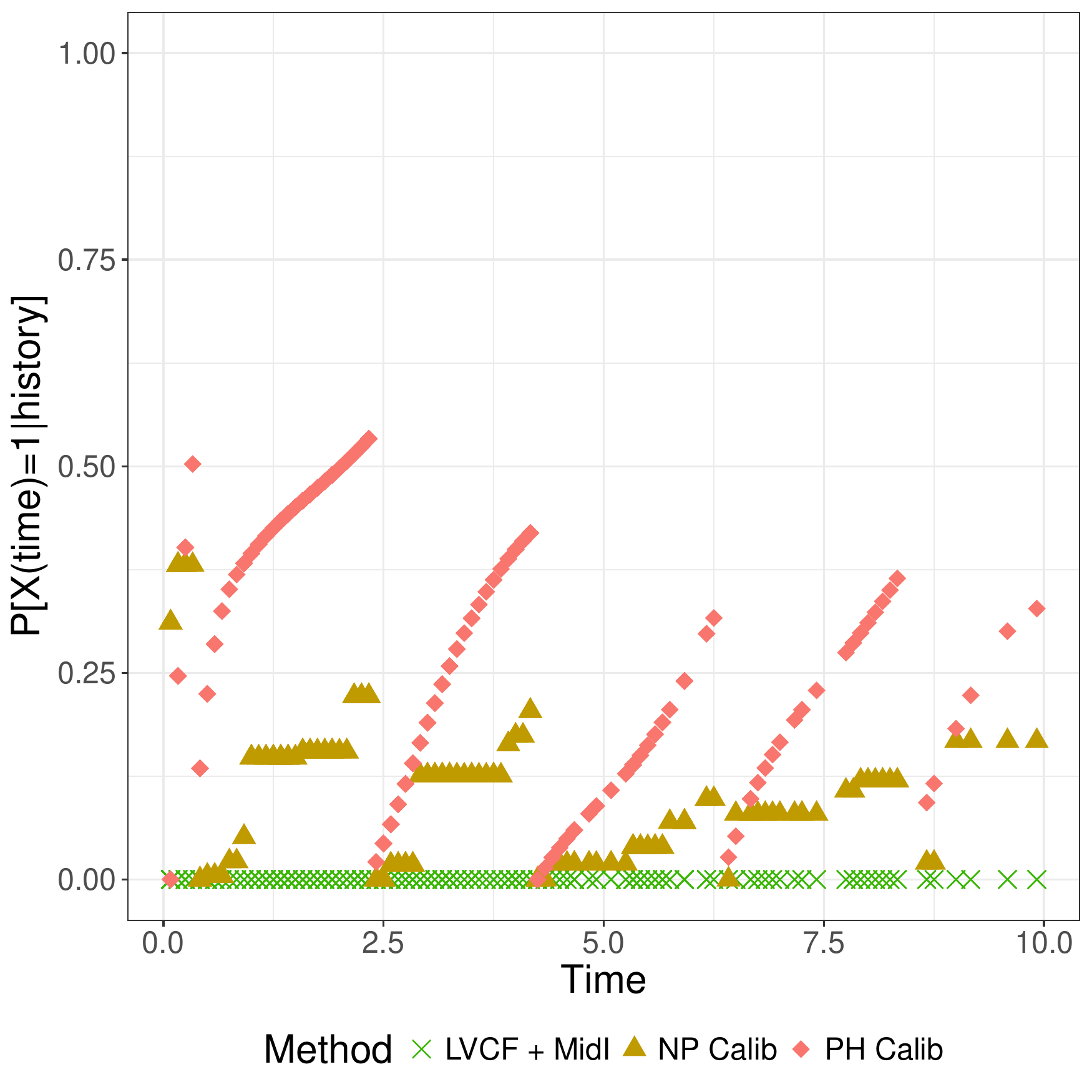}
				\caption{First participant}
				\label{SubFig:ProbsOneID}	
	\end{subfigure}
	%\hfill
	~~~~~
	\begin{subfigure}[t]{0.45\textwidth}
	\includegraphics[width=\textwidth]{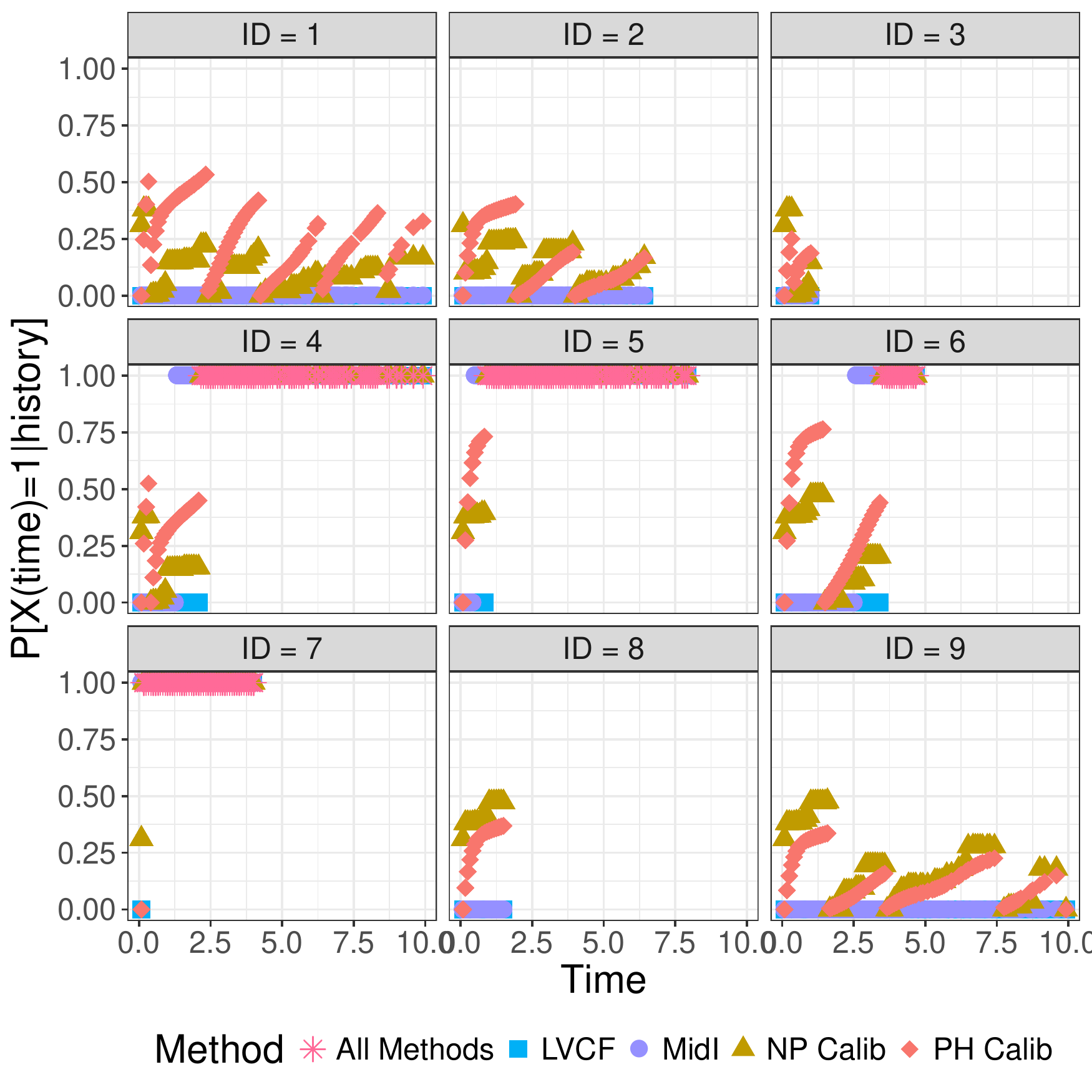}
	\caption{First 9 participants}
	\label{SubFig:ProbsNineIDs}
	\end{subfigure}
	\caption{$\widehat{P}(X_i(t)=1|\text{history})$ by the different considered methods for the first 9 participants. Panel (a) corresponds to the top left corner of Panel (b).  \label{Fig:Probs}}
\end{center}
\end{figure}

The simulation results presented in Section \ref{Sec:Sims} and the Supplementary Materials have shown that when there are covariates affecting the time-to-exposure, the PH-OC estimator is preferable over the n\"aive methods, namely LVCF and MidI. Figure \ref{Fig:Probs} illustrates the difference between the methods. On its right panel, we drew the probabilities	$\widehat{P}(X_i(t)=1|\text{history})$ versus $t$, for the first nine participants in our data, using LVCF, MidI, non-parametric (NP) calibration, and the PH calibration model.  Once $X_i(t)=1$ was observed, all methods assign $\widehat{P}(X_i(t)=1|\text{history})=1$ for the rest of the relevant risk sets.  The left panel of Figure \ref{Fig:Probs} presents the same probabilities as the right panel of the figure, but for the first participant only.  This person did not report aspirin taking during the 10 years follow-up ($\tilde{T}=10$). From the data, this person was a 75 years old (at diagnosis time) male, that was taking aspirin at baseline. From Table \ref{Tab:CalibModel}, we would expect the probability of this person to take aspirin to be higher than in the general population. This aligns with this person having larger estimated probabilities  under the PH model comparing to the NP calibration model. 

We included in the risk-set calibration models the covariates of Model (III). To avoid destabilization of the calibration model fittings, we grouped the  risk sets in intervals of size 0.5. That is, we refitted the calibration model every 6 months. 

Turning to the main model, we included  baseline covariates that are known to be associated with CRC-related death. These included  age-at-diagnosis, pre-diagnosis BMI, family history of CRC,  and the following tumor characteristics:  stage, differentiation and location. Table \ref{Tab:MainCRC} shows estimates and corresponding standard errors,  confidence intervals, and $p$-values  for the association between aspirin-taking and survival in the three motivating studies and  in the entire  data. Compared to LVCF,  stronger protective associations were estimated by our methods in all three subtype-specific analyses.  Compared to the OC estimates,  the RSC estimates were only slightly further from null. This could be partially explained by the high censoring rate ($>$80\%). Even though the sample sizes and case numbers were low to moderate,  there was evidence for strong protective effect of aspirin for PIK3CA subtype CRC. The point estimates for aspirin in low CD274 subtype imply  negative association between aspirin and death, but,  possibly due to limited power, the null hypothesis was not rejected.
 \begin{table}
 	 	\caption{\small Results for the main model in the three CRC studies and in all data. Results presented for the aspirin effect. The main model also included family history of CRC, pre-diagnosis BMI, age-at-diagnosis, gender, disease stage, differentiation and location. \label{Tab:MainCRC}}
 		 	\begin{center}
 		 		\small
 		\begin{tabular}{cccccc}
 			Study	& Method	& Est (SE) & $\widehat{HR}$  & CI 95\%(for HR)  &  $p$-value \\ 
 			\hline
 			All Data & LVCF & -0.34 (0.147) & 0.71 &   $[0.53,0.95]$ & 0.020 \\ 
 			($n=1,371$) &  PH-OC & -0.32 (0.176) & 0.73 &   $[0.51,1.03]$ & 0.072 \\  			
 			(No. Events $=249$) & PH-RSC & -0.32 (0.177) & 0.73 & $[0.51,1.03]$ & 0.073 \\ 
 			\hline
 			Low CD274$^\star$ &  LVCF & -0.64 (0.352) & 0.53 &  $[0.26,1.05]$ & 0.067 \\ 
 			($n=278$)     &   PH-OC    & -0.77 (0.400) & 0.46  & $[0.21,1.01]$ & 0.054 \\ 			
 			 (No. Events $=50$) & PH-RSC & -0.79 (0.401) & 0.45   & $[0.21,1.00]$ & 0.049 \\
 			\hline
 			PIK3CA$^\dagger$ & LVCF & -2.13 (0.639) & 0.12 &   $[0.03,0.41]$ & 0.001 \\ 
 			($n=171$)		& PH-OC &  -2.22 (0.653) & 0.11   & $[0.03,0.39]$ & 0.001 \\ 
 			 (No. Events $=28$) & PH-RSC & -2.23 (0.657) & 0.11 &   $[0.03,0.39]$ & 0.001 \\
 			\hline
 			PTGS2$^\ddagger$ & LVCF & -0.30 (0.202) & 0.74 &  $[0.50,1.10]$ & 0.138 \\ 
 			($n=672$)	  & PH-OC & -0.32 (0.241) & 0.73 &  $[0.45,1.16]$ & 0.185 \\ 
 		 	(No. Events $=125$) & PH-RSC & -0.32 (0.241) & 0.72 & $[0.45,1.16]$ & 0.181 \\ 
 			\hline
 		\end{tabular}\\
 		\end{center}
 		\footnotesize
 		$^\star$ Hamada et al. (JCO, 2017)\\
 		$^\dagger$ Liao et al. (NEJM, 2012) \\
 		$^\ddagger$ Chan et al. (JAMA, 2009)
 		\end{table}
\section{Reanalyzing the data in \cite{goggins1999applying}}
\label{Sec:CMV}
To our knowledge, the only non-n\"aive method developed for the problem of fitting a PH model with interval-censored time-to-exposure was developed by \cite{goggins1999applying}. Their motivated application was an analysis of AIDS Clinical Trial Group (ACTG) 181 \citep{finkelstein2002analysis}. The event of interest, active CMV disease, was non-terminal. Their binary covariate was CMV shedding. Based on  the joint likelihood for $\beta$ and  the distribution of $V$ ($W$ in their notation), \cite{goggins1999applying} proposed an EM algorithm where the E-step is carried out with respect to the ordering of CMV shedding among the study participants. They further developed a Gibbs sampler to improve computational efficiency. 

Our  methodology has several improvements over the method of \cite{goggins1999applying}. First,  we exploit modern and fast estimation procedures for estimating the distribution of $V$, by the NPMLE. Second,  we can (but do not have to) include parametric modeling assumptions on the distribution of $V$, if appropriate. Third, since we fit the calibration and  main models separately, we can include in our analysis participants without data about the covariate of interest.
 Finally,  and most importantly, we include measured covariates affecting the time-to-exposure $V$.

We analyzed the ACTG 181 data \citep{finkelstein2002analysis} using our method under non-parametric calibration.  The results are presented in  Table ???? in the Supplementary Materials. We observed a divergence between the OC and RSC estimates. The RSC estimates were further away from the null, and similar to those obtained by \cite{goggins1999applying}. The confidence intervals were quite wide, as one may obtain  for hazard ratios of strong effects  when the sample size is moderate.
\section{Conclusion}
\label{Sec:Discuss}
We have presented a novel calibration approach for studying the association between a time-dependent binary exposure and survival time, when the data about the monotone time-dependent exposure is only available intermittently. Our proposed approach allows for wide range of calibration models. In practice, an adequate model should be chosen by combination of subject-matter knowledge and the available data.   The \textbf{R} package \texttt{ICcalib}  implementing our methods is publicly and freely available at \textit{blinded}.

When the association between exposure and survival is strong, and a terminal event is non-rare, our calibration framework may suffer from bias, that can be reduced, though not eliminated, by risk-set calibration. Unlike regression calibration methods for error-prone covariates, additional data (e.g., reliability or validation data) is not needed to fit the calibration model, as the covariate measurements are inherently part of the data collected to answer the question of interest.

The problem described in this paper  and the proposed conceptual framework open the way for further research. A main question of interest is whether $\beta$ can be estimated consistently from the data and what is the nature of further assumptions and methods to ensure consistency. Our model is a type of a joint model for time-to-event  and longitudinal data \citep{tsiatis2004joint,rizopoulos2012joint}. Potential alternative methods may include developing  joint modeling methods for binary longitudinal covariates  \citep{faucett1998analysis,larsen2004joint, rizopoulos2008two} and adapting them to the problem of interval-censored  change-time. The bias in the calibration model can be tackled using methods for dependent interval-censoring \citep{finkelstein2002analysis}.   Another  potential way to expand the research presented in this paper is to define alternative causal estimands and investigate the identifiability assumptions needed to estimate parameters of interest.

In conclusion, we presented novel analyses to overcome a common problem in medical and epidemiological research while presenting a new conceptual framework accompanied by flexible and simple methodology  to preform time-to-event analysis under the  problem of infrequently updated binary covariate.
\bigskip
\begin{center}
	{\large\bf SUPPLEMENTARY MATERIAL}
\end{center}
Section A presents the asymptotic theory for the RSC, Section B describe our R package.  Additional simulation studies are described in Section C. Section D presents the analysis results of the CMV data.   
\bibliography{ChangePointBin}
\bibliographystyle{agsm}
\appendix
\section*{Appendix}
We first show that $\hat{\btheta}\rightarrow\btheta^\star$.
Denote $\nu_i^{\btheta,\beeta}(t)=E_{P_{\beeta}}[\exp(\beta X_i(t))|\mathcal{G}_{it}]$ and $\nu_i^{0}(t)=E_{\mathcal{P}_0}[\exp(\beta^0 X_i(t))|\mathcal{G}_{it}]$, where
 for any $\beeta$, $E_{P_{\beeta}}$ is the expectation under the PH model \eqref{Eq:PHIsplines} for $V$, and  $E_{\mathcal{P}_0}$ is the expectation under the true distribution. Denote also 
\begin{align*}
\begin{split}
\bS^{(m)}(\btheta,\beeta, t)&=\frac{1}{n}\sum\limits_{i=1}^{n}\bigg[Y_i(t)\exp(\bgamma^T\bZ_i)\nu_i^{\btheta,\beeta}(t)\ba_i(\btheta,\beeta,t)^{\otimes m}\bigg]\\
\bs^{(m)}(\btheta,\beeta,t)&=E[\bS^{(m)}(\btheta,\beeta,t)]\\
\overline{\bs}^{(m)}(\beeta,t)&=E\bigg[\frac{1}{n}\sum\limits_{i=1}^{n}\bigg(Y_i(t)\lambda_0(t)\exp(\bgamma_0^T\bZ_i)\nu_i^{0}(t)\ba_i(\btheta,\beeta,t)^{\otimes m}\bigg)\bigg]
%\tilde{\bs}^{(m)}(t)&=E\bigg[\sum\limits_{i=1}^{n}\bigg(Y_i(t)\exp(\bgamma_0^T\bZ_i)\nu_i^{0}(t)\ba_i(\btheta_0,\beeta^\star,t)^{\otimes m}\bigg)\bigg]
\end{split}
\end{align*}
where $\ba_i(\btheta,\beeta,t)=\begin{pmatrix}
\frac{\exp(\beta)P_{\beeta}[X_i(t)=1|\mathcal{G}_{it}]}{1+(\exp(\beta)-1)P_{\beeta}[X_i(t)=1|\mathcal{G}_{it}]}\\ \bZ_i
\end{pmatrix}$ and where for any vector $\bx$, ${\bx}^{\otimes0}=1, {\bx}^{\otimes1}=\bx$, and $\bx^{\otimes2}=\bx{\bx}^T$. Observe that $\lambda_0(t)\exp(\bgamma_0^T\bZ_i)\nu_i^{0}(t)$ is the true hazard function (conditionally on $\mathcal{G}_{it}$). %Note that $\tilde{\bs}^{(m=0)}(t)$ does not include $\beeta$.
%Recall that under a parametric calibration model for the distribution of $V$, indexed by a parameter vector $\beeta$, the estimating  equations for $\btheta=(\beta,\bgamma)$  are $\bU_{\btheta}(\btheta;\hat{\beeta})=0$, where
%\begin{align*}
%\begin{split}
%\label{Eq:U.theta.hat}
%\bU_{\btheta}(\btheta;\beeta)&=\frac{1}{n}\sum\limits_{i=1}^{n}\int\limits_{0}^{\tau}\left[\ba_i(\btheta,\beeta)-\frac{\bS^{(1)}(\btheta,\beeta,t)}{S^{(0)}(\btheta,\beeta,t)}\right]dN_i(t)
%\end{split}
%\end{align*}
%and where $\hat{\beeta}$ is the estimator of $\beeta$ obtained from the interval-censored data on the time-to-covariate-change. 
Define also 
$
\bu^{\mathcal{G}}_{\btheta}(\btheta;\beeta)=\int\limits_{0}^{\tau}\overline{\bs}^{(1)}(\beeta,t)dt-\int\limits_{0}^{\tau}\frac{\bs^{(1)}(\btheta,\beeta,t)}{s^{(0)}(\btheta,\beeta,t)}\overline{s}^{(0)}(\beeta,t)dt,
$  
and let  $\bI^{\mathcal{G}}_{\btheta}(\btheta,\beeta)=\nabla_{\btheta}\bu^{\mathcal{G}}_{\btheta}(\btheta;\beeta)$ and 
$\bI^{\mathcal{G}}_{\beeta}(\btheta,\beeta)=\nabla_{\beeta}\bu^{\mathcal{G}}_{\btheta}(\btheta;\beeta)$.
%\end{split}
%\end{align*}
%\begin{align*}
%\begin{split}
%\bI_\btheta(\btheta,\beeta)&=\nabla_{\btheta}\bu_{\btheta}(\btheta;\beeta)\\
%\bI_{\btheta\beeta}(\btheta,\beeta)&=\nabla_{\beeta}\bu_{\btheta}(\btheta;\beeta)\\
%\bI_\btheta(\btheta,\beeta)&=\nabla_{\beeta\btheta}\bu_{\btheta}(\btheta;\beeta).
%\end{split}
%\end{align*}
 We impose the following standard regularity assumptions:
%Define also $\bu_{\btheta}(\btheta;\beeta)$ to be the vector function $\bU(\btheta;\beeta)$ when replacing $S^{(m)}(\btheta,\beeta,t)$ with $s^{(m)}(\btheta,\beeta,t)$ defined above for $m=0,1$.  
\begin{enumerate}[label=(A\arabic*)]
	\item The number of knots $K$ does not grow with the sample size $n$. \label{Ass:knots}
	\item $\hat{\beeta}\xrightarrow{p}\beeta^\star$, for some $\beeta^\star$. \label{Ass:eta}
	\item $\bs^{(m)}(\btheta,\beeta,t), m=0,1,2$ are continuous and bounded functions of $\btheta$, for any $\btheta$ and $\beeta$ in the  neighborhoods $\Theta$ and $\mathcal{H}$ of $\btheta^\star$ and $\beeta^\star$, respectively, and for all $t\in[0,\tau]$. Furthermore, $s^{(0)}(\btheta^\star,\beeta^\star,t)$ is bounded away from zero. \label{Ass:s}
	\item The components of $\bZ_i$ are bounded for all $i$.
	\item The matrix $\bI^{\mathcal{G}}_{\btheta}(\btheta,\beeta^\star)$ is continuous in $\btheta$ and positive definite at $\btheta^\star$.  \label{Ass:Hess}
\end{enumerate}
Assumption (A1) could be relaxed, see \cite{wang2016flexible}. But for simplicity, we consider (A1) as given above.
By the Weak Law of Large Numbers,
%\begin{equation*}
$
\sup_{t\in[0,\tau],\btheta\in \Theta,\beeta \in \mathcal{H}}|\bS^{(m)}(\btheta,\beeta,t)-\bs^{(m)}(\btheta,\beeta,t)|\xrightarrow{p}0.
$
%\end{equation*}
By the assumptions above and using arguments similar to those of \cite{andersen1982cox} as implemented by \cite{lin1989robust}, it follows that for any $\btheta\in \Theta$ and $\beeta \in \mathcal{H}$, $\bU^{\mathcal{G}}_{\btheta}(\btheta;\beeta)\xrightarrow{p}\bu^{\mathcal{G}}_{\btheta}(\btheta;\beeta)$.  Recall that $\hat{\btheta}$ is the solution of $\bU^{\mathcal{G}}(\btheta;\hat{\beeta})=0$ and observe that
\begin{equation*}
\label{Eq:Ubeeta}
\bU^{\mathcal{G}}_{\btheta}(\btheta;\hat{\beeta})=\bU^{\mathcal{G}}_{\btheta}(\btheta;\beeta^\star)+(\bU^{\mathcal{G}}_{\btheta}(\btheta;\hat{\beeta})-\bU^{\mathcal{G}}_{\btheta}(\btheta;\beeta^\star))=\bU^{\mathcal{G}}_{\btheta}(\btheta;\beeta^\star)+o_p(1)
\end{equation*}
where the last equality holds by Assumption \ref{Ass:eta} and since $\ba_i$ is bounded for finite values of $\beta$. 
Let $\btheta^\star$ be the solution of $\bu^{\mathcal{G}}_{\btheta}(\btheta;\beeta^\star)=0$. By the assumptions above, and specifically Assumption \ref{Ass:Hess}, $\hat{\btheta}\xrightarrow{p}\btheta^\star$. In particular, $\hat{\beta}\xrightarrow{p}\beta^\star$.
%Finally, let  $\tilde{\ba}_i(\btheta,\beeta)=\begin{pmatrix}
%\frac{\exp(\beta)P_{\beeta}[X_i(t)=1|\mathcal{G}_{it}]}{1+(\exp(\beta)-1)P_{\beeta}[X_i(t)=1|\mathcal{G}_{it}]}\\ \bZ_i
%\end{pmatrix}$

Regrading asymptotic normally,  by a Taylor expansion 
$$
0=\bU^{\mathcal{G}}_{\btheta}(\hat{\btheta};\hat{\beeta})=\bU^{\mathcal{G}}_{\btheta}(\btheta^\star;\beeta^\star)+[\nabla_{\btheta}\bU^{\mathcal{G}}_{\btheta}(\btheta^\star;\beeta^\star)](\hat{\btheta}-\btheta^\star)+[\nabla_{\beeta}\bU^{\mathcal{G}}_{\btheta}(\btheta^\star;\beeta^\star)](\hat{\beeta}-\beeta^\star)+o_p(n^{-1/2})
$$
which can be rearranged as 
\begin{equation}
\label{Eq:betaAsyTaylor}
\sqrt{n}(\hat{\btheta}-\btheta^\star)= [-\nabla_{\btheta}\bU^{\mathcal{G}}_{\btheta}(\btheta^\star;\beeta^\star)]^{-1}\sqrt{n}\{\bU^{\mathcal{G}}_{\btheta}(\btheta^\star;\beeta^\star)+[\nabla_{\beeta}\bU^{\mathcal{G}}_{\btheta}(\btheta^\star;\beeta^\star)(\hat{\beeta}-\beeta^\star)]\}   +o_p(1).
\end{equation}
Similarly to \cite{andersen1982cox} and \cite{lin1989robust}, by the assumptions given above,   $\nabla_{\btheta}\bU^{\mathcal{G}}_{\btheta}(\btheta^\star;\beeta^\star)\xrightarrow{p}\bI^{\mathcal{G}}_{\btheta}(\btheta^\star,\beeta^\star)$, and by invoking similar arguments  $\nabla_{\beeta}\bU^{\mathcal{G}}_{\btheta}(\btheta^\star;\beeta^\star)\xrightarrow{p} \bI^{\mathcal{G}}_{\beeta}(\btheta^\star,\beeta^\star)$. Let $\mathcal{N}(t)=E[n^{-1}\sum_{i=1}^{n}N_i(t)]$. By arguments similar to those of \cite{lin1989robust}, it can be shown that $n^{1/2}\bU^{\mathcal{G}}_{\btheta}(\btheta^\star;\beeta^\star)=n^{-1/2}\sum\limits_{i=1}^{n}\bb_i(\btheta^\star;\beeta^\star)+o_p(1)$, where
\begin{align*}
\begin{split}
\bb_i(\btheta;\beeta)&=\int\limits_{0}^{\tau}\left[\ba_i(\btheta,\beeta,t) -\frac{\bs^{(1)}(\btheta,\beeta,t)}{s^{(0)}(\btheta,\beeta,t)}\right]dN_i(t)\\
&-\int_0^\infty\frac{Y_i(t)\exp({\bgamma}^T\bZ_i) \nu_i^{\btheta,\beeta}(t)}{s^{(0)}(\btheta,\beeta,t)}\left[\ba_i(\btheta,\beeta,t) -\frac{\bs^{(1)}(\btheta,\beeta,t)}{s^{(0)}(\btheta,\beeta,t)}\right]d\mathcal{N}(t)+o_p(1).
\end{split}
\end{align*}
Regarding $(\hat{\beeta}-\beeta^\star)$,  by Assumption \ref{Ass:knots}, it is a finite-size vector, and as explained in \cite{wang2016flexible}, it can be treated with the standard tools for parametric models. Denote $\ell^V=\log L^V$. By further incorporating standard theory of misspecified likelihood-based models and Assumption \ref{Ass:eta}, we may write 
\begin{equation*}
%\label{Eq:beetaAsy}
\sqrt{n}(\hat{\beeta}-\beeta^\star)= -\left(\frac{1}{n}\sum\limits_{i=1}^{n}\nabla_{\beeta\beeta}\ell^V_i(\beeta^\star)\right)^{-1}\left(\frac{1}{\sqrt{n}}\sum\limits_{i=1}^{n}\nabla_{\beeta}\ell^V_i(\beeta^\star)\right) + o_p(1).
\end{equation*}
Substituting this in \eqref{Eq:betaAsyTaylor}, we get
\begin{equation*}
\label{Eq:betaAsyTaylorSub}
\sqrt{n}(\hat{\btheta}-\btheta^\star)= [-\nabla_{\btheta}\bU^{\mathcal{G}}_{\btheta}(\btheta^\star;\beeta^\star)]^{-1}\left(\frac{1}{\sqrt{n}}\sum\limits_{i=1}^{n}\br_i(\btheta^\star,\beeta^\star)\right)   +o_p(1)
\end{equation*}
where 
$$
%r_i(\btheta,\beeta)=_i+\nabla_{\beeta}\bb_i-\left(\sum\limits_{j=1}^{n}\nabla_{\beeta\beeta}\ell^S_j(\beeta^\star)]^{-1}\right)\left(\sqrt{n}\sum\limits_{j=1}^{n}\nabla_{\beeta}\ell^S_j(\beeta^\star)\right).
\br_i(\btheta,\beeta)=\bb_i(\btheta,\beeta)-[\nabla_{\beeta}\bU^{\mathcal{G}}_{\btheta}(\btheta;\beeta)]\left(\frac{1}{n}\sum\limits_{j=1}^{n}\nabla_{\beeta\beeta}\ell^V_j(\beeta)\right)^{-1}\nabla_{\beeta}\ell^V_i(\beeta).
$$
Therefore, by the multivariate central limit theorem and Slutsky's theorem, we conclude that  $\sqrt{n}(\hat{\btheta}-\btheta^\star)$ is asymptotically normally distributed with covariance matrix 
$$
[\bI^{\mathcal{G}}_{\btheta}(\btheta^\star,\beeta^\star)]^{-1}E(\br_i(\btheta^\star,\beeta^\star)^{\otimes 2})[\bI^{\mathcal{G}}_{\btheta}(\btheta^\star,\beeta^\star)]^{-1}
$$
which can be consistently estimated by $\hat{\mathcal{V}}$ in Equation \eqref{Eq:VarEst}, by replacing parameters with their estimates and expectations with their corresponding sample version.
% $$
% [-\nabla_{\btheta}\bU_{\btheta}(\hat{\btheta};\hat{\beeta})]^{-1}\left(\frac{1}{n}\sum\limits_{i=1}^{n}\hat{\br}_i^{\otimes 2}\right)[-\nabla_{\btheta}\bU_{\btheta}(\hat{\btheta};\hat{\beeta})]^{-1}
% $$

\end{document}